\documentclass{ieeeaccess}
\usepackage{cite}
\usepackage{amsmath,amssymb,amsfonts}
\usepackage{algorithmic}
\usepackage{graphicx}
\usepackage{textcomp}
\usepackage{multirow}
\usepackage{wrapfig}

\def\BibTeX{{\rm B\kern-.05em{\sc i\kern-.025em b}\kern-.08em
    T\kern-.1667em\lower.7ex\hbox{E}\kern-.125emX}}

\begin{document}

\doi{}

\title{
 Artificial Blur Effect for Optical See-through Near-Eye Displays}

\author{
\uppercase{Shiva Sinaei}\authorrefmark{1}, 
\uppercase{Daisuke Iwai}\authorrefmark{1},~\IEEEmembership{Member, IEEE}, \\
\uppercase{and} 
\uppercase{Kosuke Sato}\authorrefmark{1},~\IEEEmembership{Member, IEEE}
}

\address[1]{Graduate School of Engineering Science, Osaka University, Toyonaka 560-8531, Japan}

\tfootnote{
}

\markboth
{Sinaei \headeretal: Artificial Blur Effect for Optical See-through Near-Eye Displays}
{Sinaei \headeretal: Artificial Blur Effect for Optical See-through Near-Eye Displays}

\corresp{Corresponding author: Shiva Sinaei (e-mail: shiasinaei98@gmail.com).}

\begin{abstract}
Saliency modulation has significant potential for various applications. In our pursuit of implementing saliency modulation for optical see-through near-eye displays, we decided to introduce a blur effect to reduce the sharpness of specific areas while preserving the sharpness of others. In this study, we used a digital micromirror device (DMD) to separate the incoming light from a scene into sharp and blurred areas. To achieve this, we integrated an electrically tunable lens (ETL), which operates in its zero optical power mode when the reflected light from the DMD represents the sharp area (i.e., the blur area is masked). Conversely, when the reflected light indicates the blur area, the ETL adjusts to non-zero optical powers. Importantly, these modulations occur at a speed that surpasses the critical flicker frequency threshold of the human eye. Furthermore, we proposed an algorithm to mitigate the artifacts around the border area between the sharp and blur areas that are caused by the magnification of the ETL. We have also developed a prototype system to demonstrate the feasibility of our method.
\end{abstract}


\titlepgskip=-21pt

\maketitle

\section{Introduction}\label{sec:introduction}
Organisms can concentrate their limited perceptual and cognitive resources on the relevant, or salient, subset of sensory data available to them. In human-computer interaction, modulation of saliency through changes in colors, size, position, etc., has proven to be a useful tool to capture attention \cite{1}. This process of changing saliency is known as saliency modulation.


Saliency modulation methods and the resulting stimuli in visual selection offer utility across various applications. Beyond serving as the foundation for further research on visual saliency detection \cite{15}, which is valuable in human cognition and behavioral sciences, saliency modulation finds application in tasks such as visual guidance \cite{2, 3, 4, 5, 8}, focus plus context visualization \cite{6, 7, 9},
full or partial diminished reality (DR) \cite{10, 11, 12, 43} and target position detection \cite{13, 14, 17}.

Augmented Reality (AR), which seamlessly blends the physical and virtual worlds, has the potential to modulate the saliency of real-world scenes in situ. Originally, saliency modulation in AR was introduced to enhance visual guidance and improve task performance \cite{4, 16, 45, 46, 48}.

AR systems are typically classified into two categories: Video see-through AR (VST-AR) and optical see-through AR (OST-AR). Saliency modulation in VST-AR can be easily achieved through the application of image and video processing techniques. Various methods have been explored to manipulate images, emphasizing certain parts over others. 
These techniques often involve color adjustments \cite{48, 18}, luminance manipulation \cite{19}, or adding a blur effect to non-target areas by convolving captured images with a varying point-spread function (PSF) map \cite{38}. Despite the advantages of VST-AR systems, such as the ability to apply image and video processing techniques, they come with inherent limitations, particularly in terms of latency and field of view (FOV).

On the other hand, OST-AR systems offer a direct way to interact with the real environment without the need to capture or model it separately. However, these systems have limitations, particularly when it comes to manipulating the environmental view. A major challenge for OST-AR researchers has been finding a solution to manipulate incoming light. For instance, effectively manipulating the incoming light from the scene to create appropriate occlusion and address the ghost-like appearance of virtual content in AR has been a major research topic in the field
\cite{20, 21, 22, 23,24, 33}.

We endeavored to develop a system capable of saliency modulation by manipulating the blurriness of different parts of the image. In recent years, because of technological progress in optics, including the accessibility of free-form lenses and high-speed electrically tunable lenses (ETL), numerous studies have leveraged these advancements to manipulate incoming light and images for various purposes. These applications range from addressing the vergence and accomodation (VAC) problem \cite{25, 33}, to the development of ophthalmology-related devices and AR systems tailored for individuals with visual disorders \cite{26,27,28}. Furthermore, these technologies have been employed for 
deblurring and depth of field (DOF) extention in both imaging and projection \cite{31, 32, 34}, as well as contributing to AR systems providing new effects, along with projection mapping \cite{29, 30}.
 
In our study, we developed an optical see-through near-eye display system (OST-NED) that supports saliency modulation through blur. This device demonstrates the feasibility of implementing the blur effect for visual augmentation using currently available technological components, overcoming the limitations previously imposed on visual augmentation by blur. In this system, a digital micromirror device (DMD) is synchronized with an ETL, which sweeps its focal range at a rate of 60 Hz—exceeding the critical flicker frequency (CFF) of the human eye \cite{42}. This synchronization provides the system with a sequence of images with varying degrees of blur. The DMD is employed to mask the incoming light, allowing us to distinguish between sharp and blur areas. 

\section{Related Work}\label{Related Work}

Vision augmentation holds a special significance in the wide range of AR system applications. AR systems have proven useful in reducing mental workload and improving task performance \cite{36, 47}. Recently, the concept of visual noise has been explored from a perspective distinct from visual augmentation. The aim is to enhance the user experience through visual noise cancellation \cite{43}.

Visual guidance, as an application of vision augmentation, can be achieved through visualization techniques, such as incorporating visual elements such as arrows, to guide users through a procedural process \cite{37}. However, the incorporation of visual elements can potentially disrupt the user experience, leading researchers to explore nuanced approaches and deploy specially designed visual elements \cite{38}. 

Another method for implementing visual guidance is saliency modulation. This technique is used in guidance systems to enhance the user's attention on essential areas by reducing the clarity of non-essential areas. Saliency modulation can be effectively achieved by deliberately blurring specific regions of a visual scene \cite{2, 3, 35}. Alternatively, this can be accomplished through the adjustment of brightness levels in non-essential areas \cite{4, 8}. or by introducing the flickering effect \cite{44, 45}.

Although the saliency modulation is suitable for the visual guidance task, implementing it in OST-AR systems poses its own challenges, as the incoming light needs to be manipulated in real-time without any control over it. Sutton et al. introduced an OST-AR system that provides saliency modulation by overlaying colorful patterns on the incoming light of the environment. In their system, a beam-splitter is used to direct a portion of light to a camera, carefully calibrated with the direction of the human eye. Based on the captured images, additive light patterns are created and overlaid onto the incoming light. Their system successfully achieved its goal \cite{4}. In our work, we aimed to implement saliency modulation using a different method besides adding additive light to the incoming environment light.

Ueda et al. \cite{30} proposed a system for saliency modulation by combining a pair of ETLs in the form of goggles with a high-speed projector. In their system, the high-speed projector is synchronized with the ETLs. If the target object is intended to appear sharp, it is illuminated by the projector when the optical power of the ETL is zero. Conversely, if the object is meant to appear blurry, it is illuminated when the optical power is non-zero. Our approach shares similarities with Ueda et al. \cite{30}, particularly in the use of ETLs to manipulate the blurriness of incoming light. However, our system employs a DMD to differentiate between blurred and non-blurred areas. In contrast, \cite{30} relied on a high-speed projector for this distinction, which restricts its applicability to dark environments. Moreover, the use of a projector makes their system non-portable. Our main contribution over this previous work is overcoming these limitations by eliminating the need for a high-speed projector. We achieve this by applying a light-blocking mechanism-essentially the reverse of the spatial illumination concept in~\cite{30}-to separate blurred and non-blurred areas.

\begin{figure*}[t]
    \centering

    \includegraphics[width=0.98\hsize]{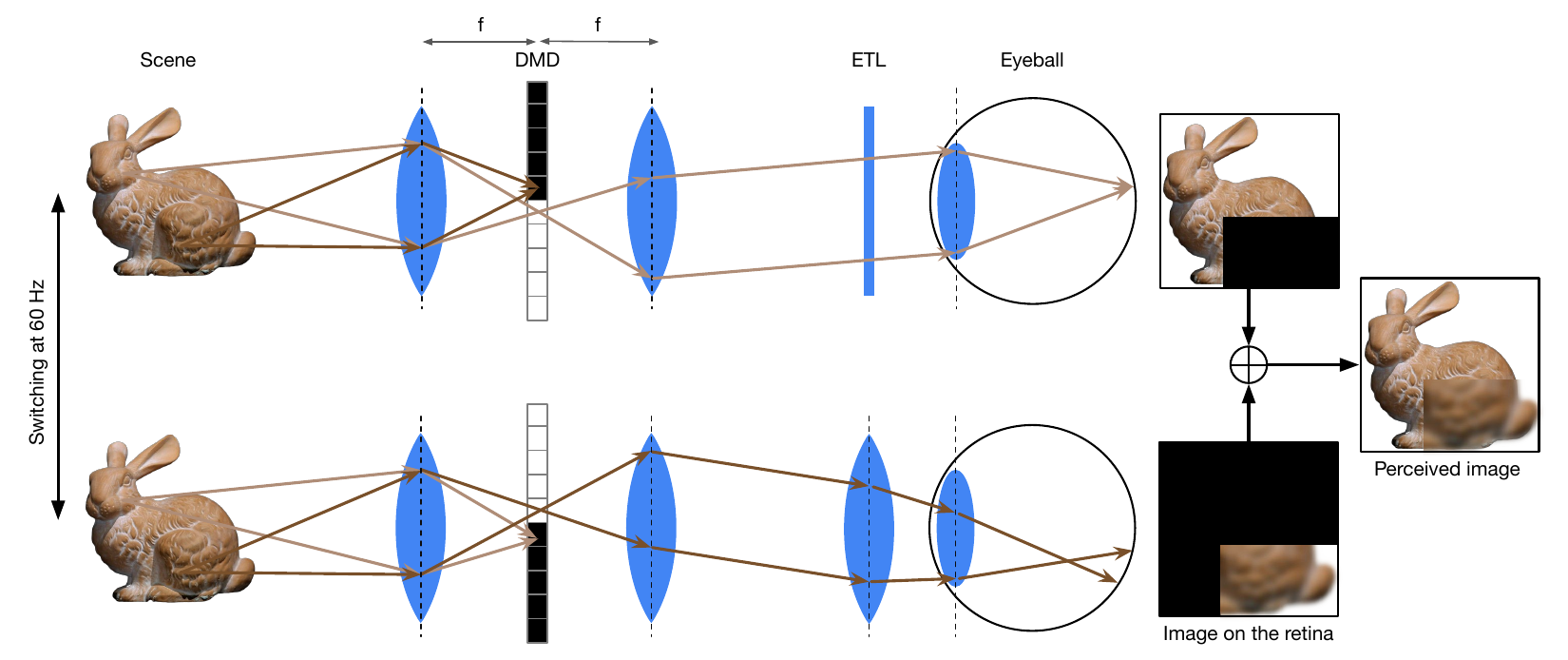} 
    \caption{Concept of the system: Some rays are obstructed according to the DMD pattern. Unobstructed rays pass through the ETL. The upper part of the figure illustrates the passage of rays through the ETL when the optical power is nearly zero (sharp image), while the lower part depicts the passage when it exceeds zero diopter (blur image). This cycle of ray traversal through the ETL occurs continuously at a frequency of 60Hz. Note that the two lenses sandwiching the DMD are used to focus the incoming light rays from the scene onto the DMD for spatial blocking, ensuring that the FOV of the blocked scene matches that of the original scene.}
    \label{fig:1}
\end{figure*}

\section{Method} 
In this section, we will explore the technical aspects and optical foundations of our system. First, we will explain how we separate the area into two parts: blur and sharp areas and how the blur is generated. Then, we will discuss our synchronization module and the signals necessary to synchronize the ETL and DMD. In the final subsection, we discuss our proposed algorithm to mitigate the artifacts that occur around the border areas between sharp and blurry regions due to the changing magnification by the ETL.

The core concept of our system involves guiding the incoming light from the scene into the DMD, which operates in two modes: (1) blur mask and (2) non-blur mask. This allows us to discern the incoming light into two distinct areas: a blur area and a sharp area. To achieve the blur effect, we utilize an ETL. When the reflected light from the DMD needs to remain sharp, the optical power of the ETL is set to zero or near zero. On the other hand, when blur is required, the ETL's optical power is adjusted to be greater than that for the sharp area, thereby refracting the light rays at sharper angles and causing blur in the image. The concept is thoroughly illustrated in Fig. \ref{fig:1}.

To guide the light through this process without distorting the FOV of the system, we apply the optical relay principle outlined in \cite{39}\cite{40}, which will be explained in more detail.

\subsection{Light Guide and Image Formation}
In the proposed OST-NED system, our objective is to manipulate the incoming light to the eye in such a way that certain areas, non-focused area, are blurred while others, focused-area remain unaltered. A critical aspect of OST-NED systems is ensuring that the incoming light from the environment, particularly the areas designated not to be overlaid or manipulated with an image or effect, remains unaffected.

In our approach, we utilize a DMD to divide the incoming light into two distinct areas. The DMD is a display device composed of thousands of micromirrors that rotate around their diagonal axis. These micromirrors switch between on and off states, with the "on" mode directing light to the desired direction and the "off" mode redirecting it elsewhere, in our system, towards a black absorbing material. The main challenge is to guide the light to the DMD and then to the eye without causing any shifts or distortions. 

\begin{wrapfigure}{r}{4cm} 
    \centering
    \includegraphics[width=4cm]{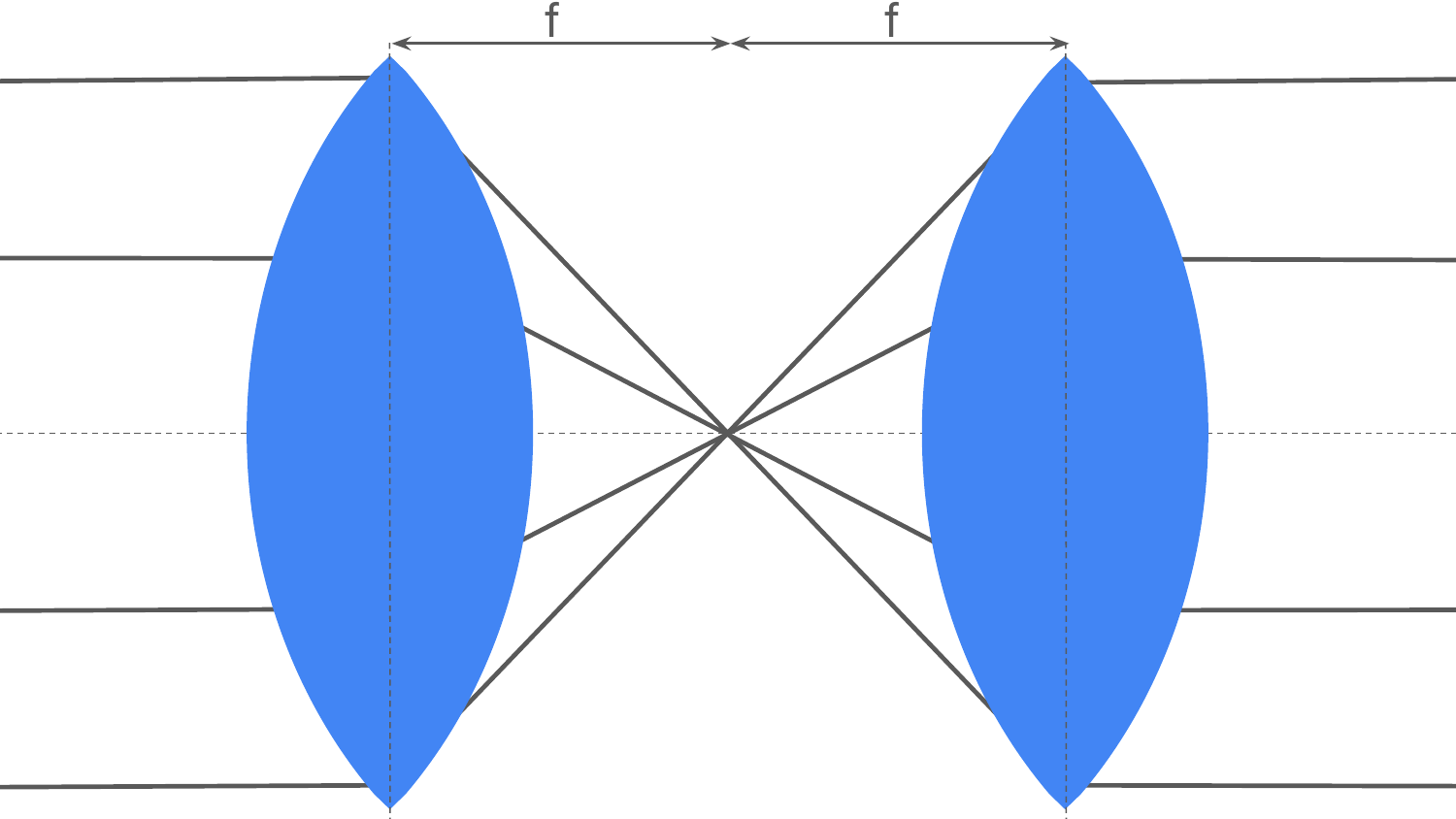}
    \caption{2f relay system}
    \label{fig:2f_relay}
\end{wrapfigure}

As proposed in \cite{39}\cite{40}, to mask the light without distortion or shift, the mask should be placed at the center of the 2f relay system. A 2f relay system consists of two lenses placed at a distance of 2f from each other. The first lens focuses the parallel incoming light at its focal length, and the light becomes parallel again after passing through the second lens (Figure \ref{fig:2f_relay}).
\vspace{-0.2cm}
\vspace{-0.2cm}

Utilizing this feature of the 2f relay system, we position the DMD at the center of the 2f system to mask the focused and non-focused areas.

After traveling through the optical relay path, the incoming light from the scene undergoes manipulation by the ETL just before it reaches the human eye. This manipulation creates blur when the light reflected by the DMD corresponds to the non-focused portion, ensuring that the desired areas are appropriately blurred by  higher optical power of the ETL. If the reflected light corresponds to the focused area, the ETL's optical power is set to zero or near zero, allowing light to pass without additional blur.

The incoming light to the ETL is collimated, meaning it will be focused exactly on the focal length of the lens when passing through it. For simplicity, we consider that the human eye is focused at infinity and has a static optical power. Consequently, when the ETL is not applying any manipulation, the incoming light creates a sharp image of infinity on the human retina. On the other hand, we know that adding a lens to an optical setup, in this case the human eye, can shift the area that the optical setup is focusing on accordingly. Therefore, when the ETL's optical power is greater than zero, the area imaged on the retina will shift from infinity to a closer area, causing the eye, which is focused on infinity, to perceive it as blurry.

Assuming the human eye has a CFF of 60 Hz, although this number depends on various factors such as the brightness of the environment light, meaning that if a sequence of images is presented to the eye at this frequency, they will be fused together, appearing as a single cohesive image. Essentially, the eye integrates these individual frames. In our setup, as previously mentioned, we divide the incoming light into two distinct areas using the DMD and then pass each area through the ETL under different optical powers. For sharp areas, the optical power should be set to zero, whereas for blur areas, a higher optical power is necessary to be applied. When these manipulated images are presented to the human eye at a speed exceeding the CFF, the eye perceives them as a unified image, incorporating both the blurred and sharp regions seamlessly. This capability is enabled by recent advancements in high-speed ETL capable of rapidly sweeping through different optical powers and displays such as DMDs, which can swiftly alter their patterns.

\subsection{Time-multiplexing with Synchronized DMD and the ETL}
Achieving saliency modulation through a blur effect and manipulating incoming light to avoid noticeable delays requires the use of high-speed devices and precise synchronization between them. As discussed previously, in our system, we divide the incoming light into two parts: focused and non-focused, in a time-multiplexing manner, making the maintenance of intensity balance between these areas crucial. While reducing the intensity of the non-targeted area, or making those areas darker, is a typical approach for saliency modulation, our objective of preserving contextual information in the non-focused area necessitates a balanced strategy. Therefore, It is essential that the overall time duration during which the focused area's light is returned to the eye not significantly differ from that of the non-focused area within one period, ensuring that information in the non-focused areas is not completely obscured. 

Additionally, the ETL lens operates using electro-wetting technology, where changes in the driving current adjust the optical power by altering the shape of the optical fluid molecules. Ideally, in order to achieve the brightest possible image with balanced brightness between the non-blurred and blurred areas, while also providing a recognizable amount of blur, a 60 Hz periodic square signal with a customized duty cycle would be optimal. However, the ETL's response time and settling time are affected by the rippling of the optical fluid and its high-frequency sinusoidal resonance until stabilization \cite{41}. Consequently, obtaining an ideal square-shaped optical power wave in response to a square current input signal is not feasible. Therefore, we opt for a sinusoidal current signal, as its optical power changes are more moderate compared to the square signal counterpart. We have customized this sinusoidal driving current so that the achieved optical power wave is always in the positive domain, in other words, functioning as a convex lens.

Given these implementation limitations, we must make a trade-off between the overall image brightness, achieving precisely zero optical power in the non-blurred area, and maintaining brightness balance between the blur and non-blur areas. Our preliminary experiments indicate that the duration during which we can achieve zero diopter optical power in response to a sinusoidal current signal to the ETL is very short. Consequently, insisting on using exact zero diopter power results in a very dark overall image. Based on the circle of confusion of the human eye, studies show that there is a range of optical powers where the bends applied to parallel rays are not that strong to be detected by the human eye \cite{30}. Therefore, it is possible to define an acceptable optical power range for the non-blurred area as ranging between \(d_{L0}\) and \(d_{L1}\) diopters (\(d_{L0}\)$<$\(d_{L1}\)). Defining \(\Delta t_{L}\) as the duration during which the optical power ranges from \(d_{L1}\) to \(d_{L0}\) and again to \(d_{L1}\) diopters (considering the sinusoidal nature of the optical power sweep) in one period of a 60Hz sinusoidal signal, in order to maintain brightness balance, it is necessary to choose the acceptable optical power range for the blurred area to be within \(d_{H0}\) to \(d_{H1}\) diopters  (\(d_{H0}\)$<$\(d_{H1}\)), such that the duration of the optical power wave from  \(d_{H0}\) to  \(d_{H1}\) and again to  \(d_{H0}\), \(\Delta t_{H}\), is nearly equal to \(\Delta t_{L}\) in duration of one period \(\Delta t\) (Fig. \ref{fig:3}). During the remaining time of the period, \(\Delta t_{B} = \Delta t - (\Delta t_{H} + \Delta t_L)\), we use a black mask to block all incoming light from the environment and preserving light balance. The patterns on the DMD, including the blur mask for reflecting the blur area, the non-blur mask for reflecting the non-blur area, and the black mask for adjusting the brightness between the non-blur area and the blur area, should be refreshed accordingly. 

\subsection{Addressing Magnification Effects at Border Regions Between Non-Blur and Blur Areas}
\begin{figure}[t]
    \centering
   \includegraphics[width=0.98\hsize]{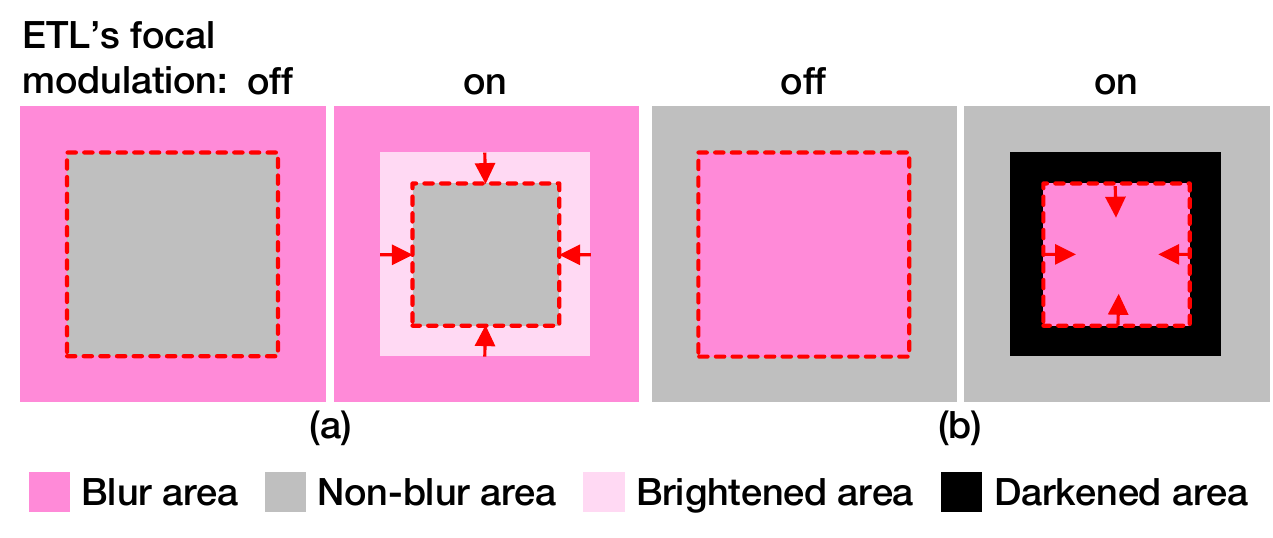} 
    \caption{ Border artifact due to magnification. (a) Artifacts in case of center in focus, (b) Artifacts in case of center out of focus.}
    \label{fig:Border}
\end{figure}

Changes in optical power within an optical system not only lead to axial image shift, resulting in blur, but also induce a magnification effect. According to the principles of geometric optics, if the object distance remains constant, the relative magnification resulting from changes in optical power can be easily quantified by dividing the image distance before and after the alteration in optical power. As previously mentioned, we have selected a range of positive optical powers, signifying the ETL's function as a convex lens. Consequently, the magnification factor always falls between 0 and 1, 
and inverted.
This suggests that as the lens's optical power increases, the resultant image will undergo greater condensation around the optical axis. Consequently, if a non-blurred area is encompassed by a blurred area, some pixels from the blurred region will leak into the border region of the sharp area, causing those pixels to appear brighter (Fig. \ref{fig:Border}-a). Conversely, when a blurred region is surrounded by a sharp area due to shrinkage, a black region will appear between these two areas (Fig. \ref{fig:Border}-b). To address the artifact of bright and dark areas, we propose a method that involves adjusting the black areas (unreflected parts) and white areas (reflected parts) of the mask, while considering the average optical power over a given duration. 

\begin{figure}[t]
    \centering
    \includegraphics[width=0.98\hsize]{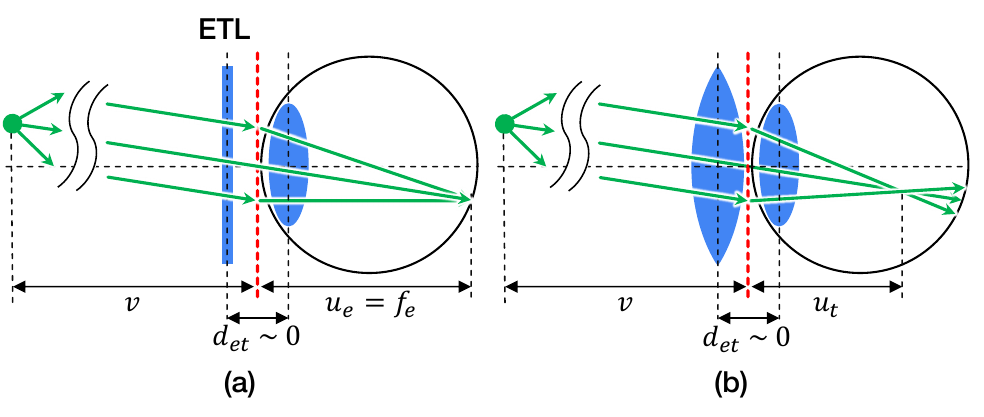} 
    \caption{ ETL-eye two-lens: Image of one point in infinity (a) when the optical power of the ETL is zero diopter; the magnification caused by the ETL is one, (b)  when optical power is greater than zero; the magnification ranges between zero and one. The dotted arrow in (b) shows that when optical power is greater than zero, the image of point on human retina is blurry. Red dashed line is the principle plane for ETL-eye lens.}
    \label{fig:Magnification}
\end{figure}

As mentioned in section III-A, the light from the scene is directed towards the DMD and then reflected back into our optical system. It eventually reaches the \textit{ETL-eye} lens, which is a two-lens optical system consisting of the ETL and the human eye lens. These two lenses are positioned very close to each other, at a distance close to zero. We define the overall optical power as $P_o$, the optical power of the human eye as $P_e$, the optical power of the ETL as $P_t$, and the distance between the eye and the ETL as $d_{et}$ (Fig. \ref{fig:Magnification}). Optical power is the inverse of the focal length of an optical element. Using the lens formula for two-lens optical systems, we can express this relationship as follows:\\
\begin{equation}
P_o = P_e + P_t - {d_{et} \cdot {P_e \cdot P_t}}
\label{eq:twolens_formula}
\end{equation}

From Eq. \ref{eq:twolens_formula} and under the assumption that the human eye and ETL are positioned at a distance of zero from each other, we can conclude that the optical power of the \textit{ETL-eye} lens is the sum of the optical powers of each lens. Considering $v$ representing the object distance, $u_e$ representing the image distance when the optical power of the ETL is zero, in other words when the eye lens is the only effective lens in system, and $u_t$ representing the image distance when the ETL has a non-zero optical power, the case when both ETL and the eye lens are active and manipulate the incoming light, we can calculate $P_e$ based on the lens formula as follows:\\
\begin{equation}
P_e = \frac{1}{v} + \frac{1}{u_e}
\label{eq:lens_formula}
\end{equation}
Similarly, for the case when both the eye lens and the ETL are active, we can write:
\begin{equation}
P_e + P_t = \frac{1}{v} + \frac{1}{u_t}
\label{eq:lens_formula2}
\end{equation}

Based on Eq. (\ref{eq:lens_formula}) and Eq. (\ref{eq:lens_formula2}), we can write the summation of $P_e$ and $P_t$ which is our overall optical power $P_o$ as follows :\\
\begin{equation}
\frac{1}{v} + \frac{1}{u_e} + P_t = \frac{1}{v} + \frac{1}{u_t}
\label{eq:lens_formula3}
\end{equation}
consequently,\\
\begin{equation}
P_t = \frac{1}{u_t}- \frac{1}{u_e} 
\label{eq:lens_formula4}
\end{equation}
Based on the definition of relative magnification as  $M = \frac{u_t}{u_e}$ and Eq. (\ref{eq:lens_formula4}), the magnification can be calculated in the following manner:\\

\begin{equation}
\frac{u_t}{u_e} = \frac{1} {1+u_e \cdot P_t }
\label{eq:lens_formula5}
\end{equation}\\
Considering $u_e$ is equal to the focal length of the human eye and is constant, and the fact that we have values of $P_t$ at different times, we can calculate the relative magnification. In our system, the incoming light on the ETL-eye side is the light that is reflected by the DMD. The FOV of the system is defined by the size of the DMD. Therefore, we can easily calculate the number of pixels shifted in each direction due to magnification, for optical power $P_t$, as follows:\\

\begin{equation}
\begin{bmatrix}
s_h(P_t)\\
s_w(P_t)
\end{bmatrix} = 
\begin{bmatrix} d_h \\ d_w \end{bmatrix} - \left( \frac{1}{1 + u_e \cdot P_t} \right) \begin{bmatrix} d_h \\ d_w \end{bmatrix}
\label{eq:shift}
\end{equation}\\

In Eq. (\ref{eq:shift}), $d_w$, $d_h$, $s_w$, and $s_h$ respectively represent the width and height of the DMD, as well as the number of pixels shifted in the width and height directions. 

By identifying the number of border pixels affected by magnification for each optical power, we can prevent light intrusion from surrounding pixels due to shifts. This can be achieved by adjusting the pixels in the blur area near the non-blur area. When there is a shift in pixels, with a width of $s_w (P_t)$ and a height of $s_h (P_t)$, we black out the closest $s_w (P_t)$ by $s_h (P_t)$ pixels in the blur area adjacent to the non-blur area to prevent light intrusion. On the other hand, if a black border appears due to image shrinkage, we compensate by turning the nearest $s_w (P_t)$ by $s_h (P_t)$ pixels in the non-blur area to white in the blur mask. 

\begin{figure}[t]
    \centering
  
    \includegraphics[width=0.98\hsize]{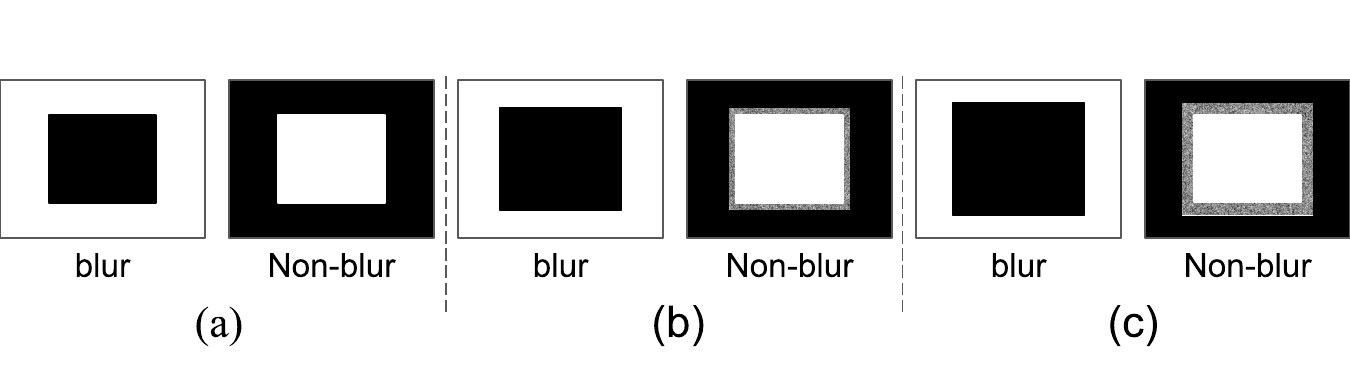}
    \caption{ An example of masks used on DMD for the case of center region in focus, with \(d_{H0}\) to  \(d_{H1}\) divided into two optical power levels P1 and P2 . (a) simple blur and non-blur mask, (b) border-compensating blur mask for optical power level 1, the $s_w (P_t)$ by $s_h (P_t)$ pixels in the blur area adjacent to the non-blur area(center) are blackened and border-compensating non-blur mask, the complementary of (c) with added random pattern in border area for optical power P1, (c) border compensating blur and non-blur mask for optical power level P2.
     }
    \label{fig:masks}
\end{figure}

The number of shifted pixels on the DMD depends on the changing optical power of the ETL during a sweep and the size of the DMD. While the optical power sweep is continuous, the DMD is a pixel-based device and thus discrete. Therefore, we need to divide the continuous range of optical powers between \(d_{H0}\) and \(d_{H1}\) into \(N\) levels. To calculate the shifted pixels $s_w (P_t)$ and $s_h (P_t)$ at each level, we take into account the average optical power of each level. We refer to these masks, which consider these pixels, as border-compensating blur masks. Figure \ref{fig:masks} showcases an example of masks designed for the center area in focus. 

As expected, the magnification effect causes some pixels to shift into the blackened or remained in whitened border area, but these shifted pixels are affected by defocus blur. Consequently, the overall intensity in these border areas is made up of overlapping distributed intensity of these blurred pixels from one side, creating a smeared pattern with different brightness compared to the non-blur and blur regions. To address this difference in light intensity within the border area, we employ a combination of masks during the non-blur period, namely border-compensating non-blur masks. During this time, we alternate between a simple non-blur mask and a series of border-compensating non-blur masks, created by calculating the complementary masks from the border-compensating blur masks. This time, however, we introduce dithering to the designated border area,  by adding a uniform binary random pattern, resulting in smoother transitions from non-blur to blur regions (Figure \ref{fig:masks}-(b, c)). The use of dithering in the border area offers an alternative to precise quantitative measurements of defocus blur, which can be computationally intensive. Introducing randomness ensures a more uniform filling of the border area, minimizing visible artifacts and enhancing the overall light transition between blurred and non-blurred regions.

\section{Implementation and Results}
In this section, we will begin by detailing our hardware and optical setup. Following this, we will explain the implementation consideration in our synchronization and border artifact alleviation employed. Subsequently, we will present the results from the perspective of the human eye to validate our approach.

\subsection{Hardware and Optical Setup}

The main components of our system consist of a DMD and an ETL. For our setup, we employed DLP5500 (0.55" XGA) DMD from Texas Instruments. To control the DMD, we utilized DLPC900 controller to change the spatial masks.

The ETL employed in our system is Optotune EL-16-40-TC. It possesses a 16mm aperture and a maximum frequency of 60 Hz. To provide the necessary input current signal to the ETL for sweeping the optical powers and to supply the trigger signals required for synchronization between the DMD and the ETL, we utilized the National Instrument USB-6343 Digital Analog Converter (DAC) with 16-bit resolution. Additionally, as the ETL necessitates current input while the output of the DAC is voltage, we utilized a customized operational amplifier specifically designed for the ETL.

\begin{figure}[t]
    \centering
    \includegraphics[width=0.98\hsize]{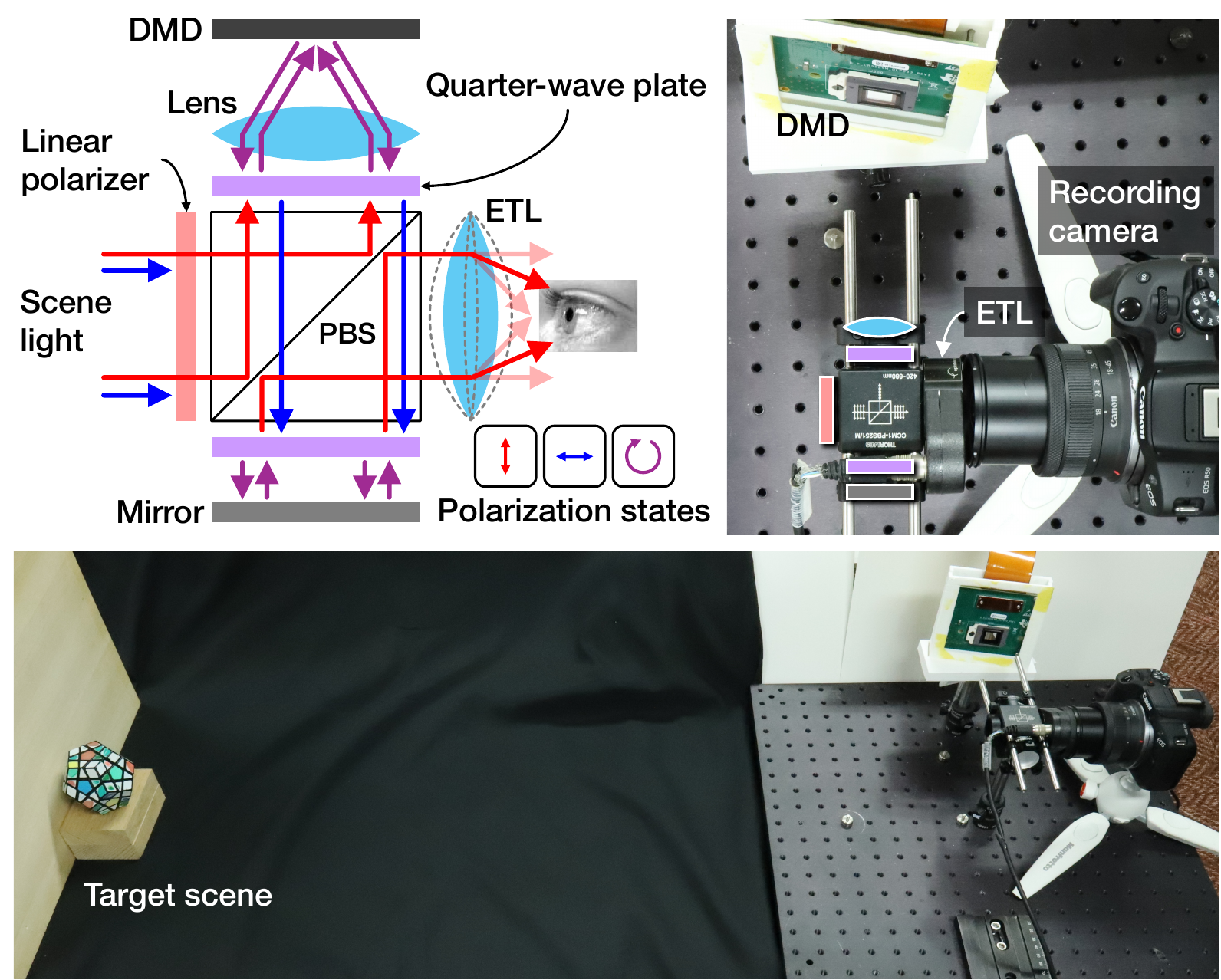} 
    \caption{Proposed system: (a) conceptual diagram, (b,c) implemented experimental setup.}
    \label{fig:6}
\end{figure}

For guiding the light to the DMD and then toward the ETL, eventually reaching the eye, our optical relay path includes a series of linear polarizers, circular polarizers, a polarizing beam splitter (PBS) and a mirror. we employed a 100 mm achromatic doublet (AC254-100-A-ML) for the 2f setup in our system. We also incorporated linear polarizers (LPVISE100-A), quarter-wave plate polarizers (WPQ10E-546-$\phi$1), and a PBS cage cube (CCM1-PBS251/M 30 mm). The relay path of light to the human eye and the prototype of our system are illustrated in Figure \ref{fig:6}-(a) and Figures \ref{fig:6}-(b,c) respectively.

\subsection{Synchronization between DMD and ETL}
\begin{figure}[t]
    \centering
    \includegraphics[width=0.98\hsize]{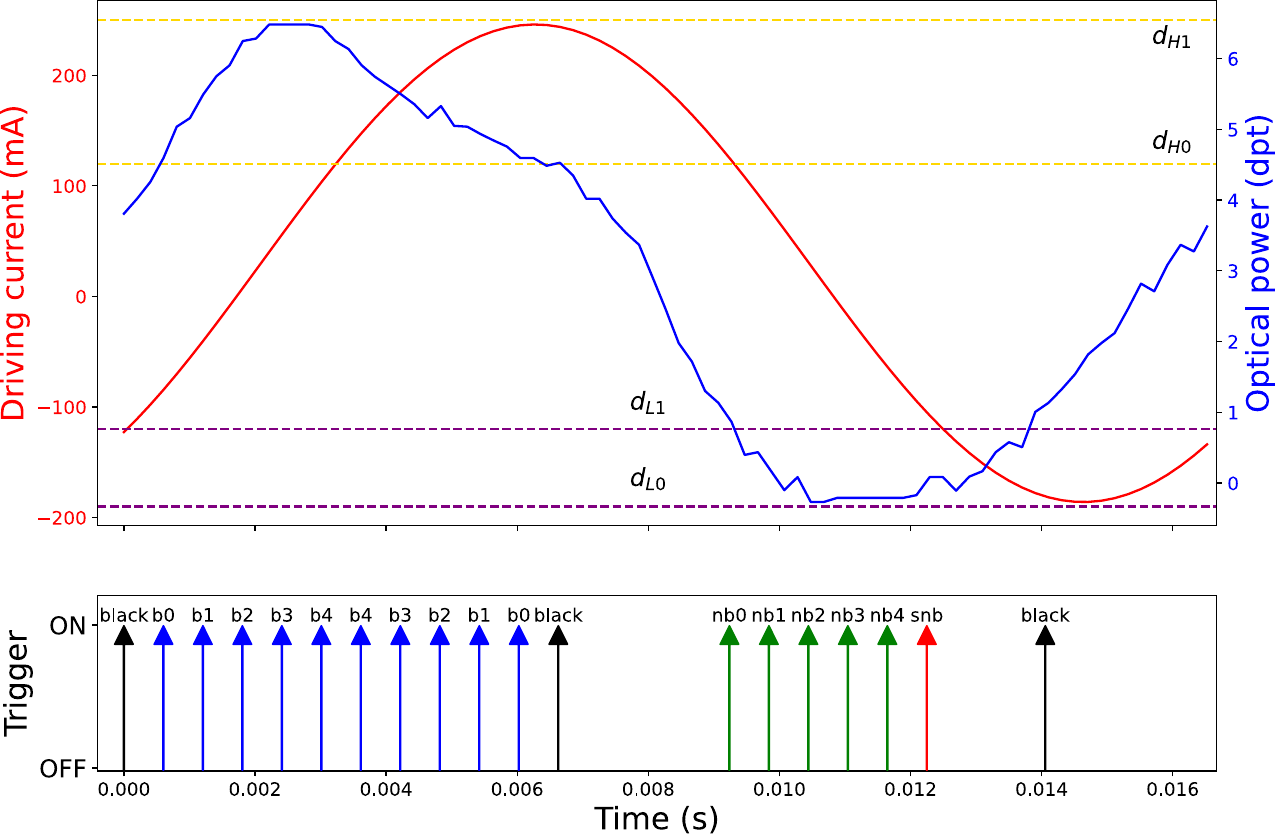} 
    \caption{The input sinusoidal current at a frequency of 60Hz (Red) and the optical power response of the ETL to this signal (Blue). Yellow lines show the \(d_{H0}\) and \(d_{H1}\), while purple lines show the \(d_{L0}\) and \(d_{L1}\). The bottom row shows trigger signals for changing the pattern on DMD within 60Hz period. Black trigger (black mask), blue trigger (border-compensated blur patterns), green trigger (border-compensated non-blur patterns), red trigger (simple non-blur pattern)}
    \label{fig:3}
\end{figure}

As discussed earlier, we have decided to use a customized sinusoidal input current for our ETL to ensure that the resulting output optical power is always greater than or equal to zero. Through our preliminary experiments, we observed the behavior of the ETL's optical power in response to a sinusoidal current input of 60 Hz. This behavior follows a sinusoidal pattern similar to the input current, but with a temporal phase shift. The response wave includes both positive optical powers, creating a convex lens effect, and negative optical powers, creating a concave lens effect. To increase the level of blurriness and make the compensation for artifacts caused by the lens magnification easier, we tried applying an offset to the input sinusoidal current. This offset ensures that the optical powers are always positive or zero. The amount of shift and offset has been calculated based on the experiments. Our experimental driving current signal to the ETL is in the form of a sinusoidal wave with an amplitude of 240 mA, a phase shift of -0.25 $\pi$, and an offset of 30 mA. This input current provides optical powers between zero and 6.5 diopters. It is worth noting that these values should be adjusted depending on the ETL model and the desired level of blurriness. The sinusoidal input signal and the resulting optical power of the ETL can be seen in Fig. \ref{fig:3}.
\\

As discussed in Section III-B, it is important to ensure consistent brightness across areas of sharpness and blurriness. Therefore, we have selected -0.1 and 0.8 diopters for \(d_{L0}\) and \(d_{L1}\) respectively, based on our preliminary experiments. Using these values, the estimated \(\Delta t_{L}\) would be approximately 5.22 ms. As a result, to achieve an approximately equal \(\Delta t_{H}\) to \(\Delta t_{L}\), we have determined that \(d_{H0}\) should be set at 4.5 diopters and \(d_{H0}\) at 6.5 diopters (\(\Delta t_{H}\) = 6.02 ms). During \(\Delta t_{n}\), which is estimated to be around 5.4 ms using the aforementioned values, we employ black masks on the DMD.

Based on these considerations, the input trigger signal implemented on the DMD will resemble what is shown in Fig. \ref{fig:3}. In this visualization, the black trigger signals correspond to the black masks, the green signals represent the border-compensating non-blur mask, the red signals indicate the simple non-blur mask, and the blue signals denote the border-compensating blur masks.

\subsection{Border Region Artifact Alleviation}

Given the pixel-based nature of the DMD display and its limited refresh rate within a 60Hz duration, it's necessary to discretize the sweep range of our blur maker optical powers, between \(d_{H0}\) and \(d_{H1}\), into $n$ different levels. The choice of $n$ depends on factors such as the DMD size, maximum refresh rate, and maximum optical power of the ETL. A greater number of masks could result in smoother transitions between non-blur and blur areas. In our implementation of border compensation, we have chosen $N$ to be 5.

For each level, we have defined border-compensating non-blur masks, namely $nb0$, $nb1$, $nb2$, $nb3$, and $nb4$, and border-compensating blur masks, namely $b0$, $b1$, $b2$, $b3$, and $b4$. Additionally, we utilized a black mask to balance the overall image brightness and a simple non-blur mask $snb$ for only reflecting the non-blur area without any border considerations, the white area of this mask is our predetermined in-focus non-blur area.

Furthermore, as discussed in Section III-C, to address the discrepancy in brightness levels caused by magnification and defocus blur effects, we introduced a dithering effect by incorporating a uniform random pattern of white and black into the border area for each border-compensating non-blur mask. The duration of blur masks for each optical power in the border-compensating blur mask and their corresponding masks in the non-blur area were set to twice as long.

\subsection{Results}

\begin{figure}[t]
    \centering
    \includegraphics[width=0.98\hsize]{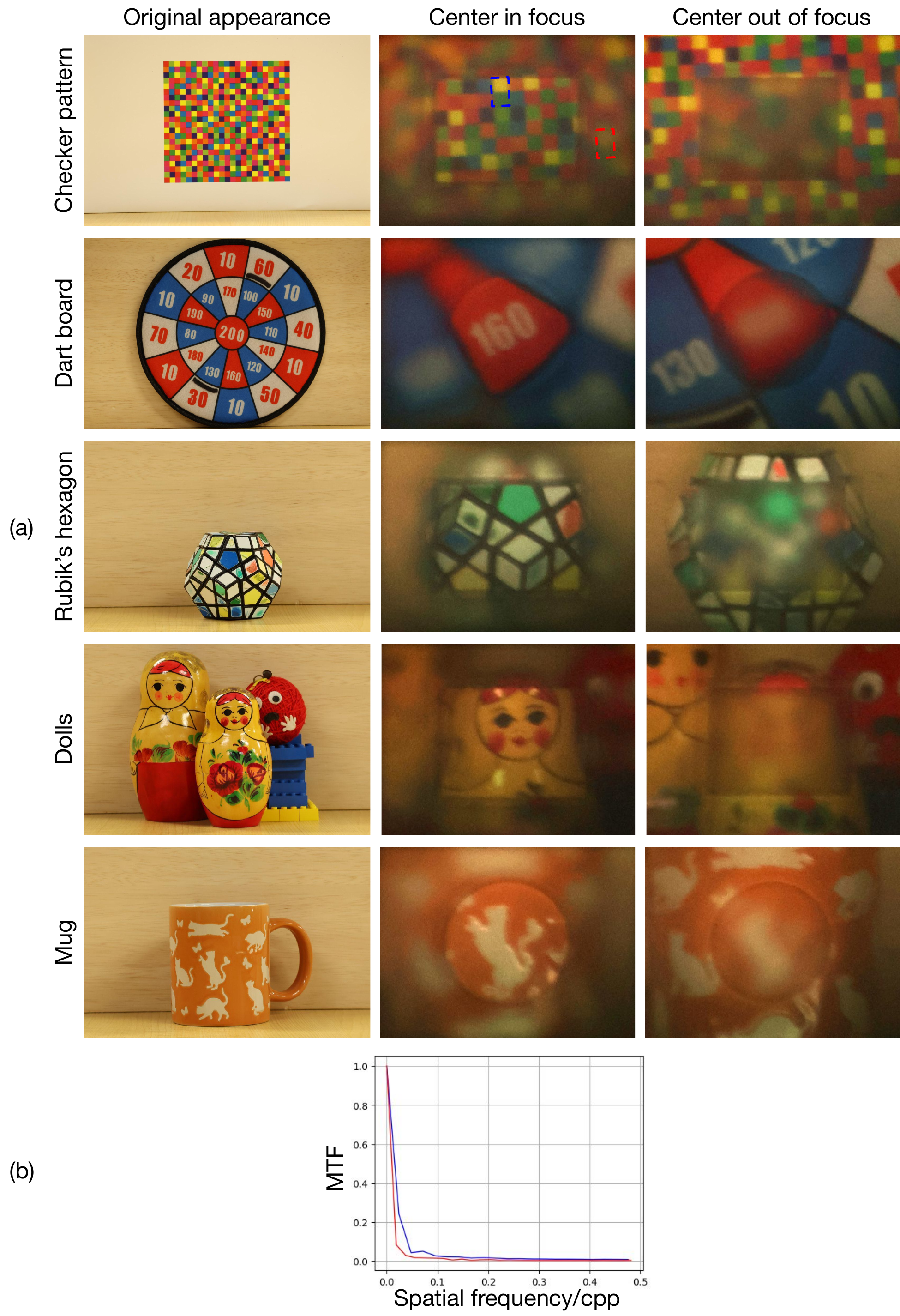} 
    \caption{Results of the system for a colorful checker pattern, dartboard, Rubik's hexagon, dolls, and a mug (a). The MTF diagram evaluated for non-blurred (blue) and blurred (red) regions (b).}
    \label{fig:allimages}
\end{figure}

To validate the effectiveness of our system in generating optical blur, we present images captured from the target object after passing through our optical setup. For image capture, we used a Canon EOS R50 camera with an ISO setting of 1250, paired with an 18 mm lens with a numerical aperture of F/4.5. The camera was positioned to simulate the perspective of a human eye. We demonstrate results from five different targets: a colorful checker pattern, a dartboard, a Rubik's hexagon, a mug, and dolls. Each target is shown twice—once with the center area in focus and once with the center area out of focus, while the surrounding area remains in focus (Fig. \ref{fig:allimages}-a). The Modulation Transfer Function (MTF) was evaluated for both the non-blurred and blurred regions, indicated by blue and red boxes, respectively. The MTF in the blurred region decreases more rapidly and remains lower across nearly all spatial frequencies compared to the non-blurred region, clearly illustrating the effect of blur.

It should be mentioned that due to the inversion caused by the lenses and mirror used in the system, the captured images are rotated by 180 degrees. However, this issue can be easily resolved by adding one more lens in front of the mirror and a 2f arrangement. Moreover, as discussed in the subsequent sections, the polarizers will affect the brightness of the incoming light, resulting in dark captured images. To overcome this, we positioned a light source in front of our target object to increase the brightness of the scene.

\begin{figure}[t]
   \centering
    \includegraphics[width=0.98\hsize]{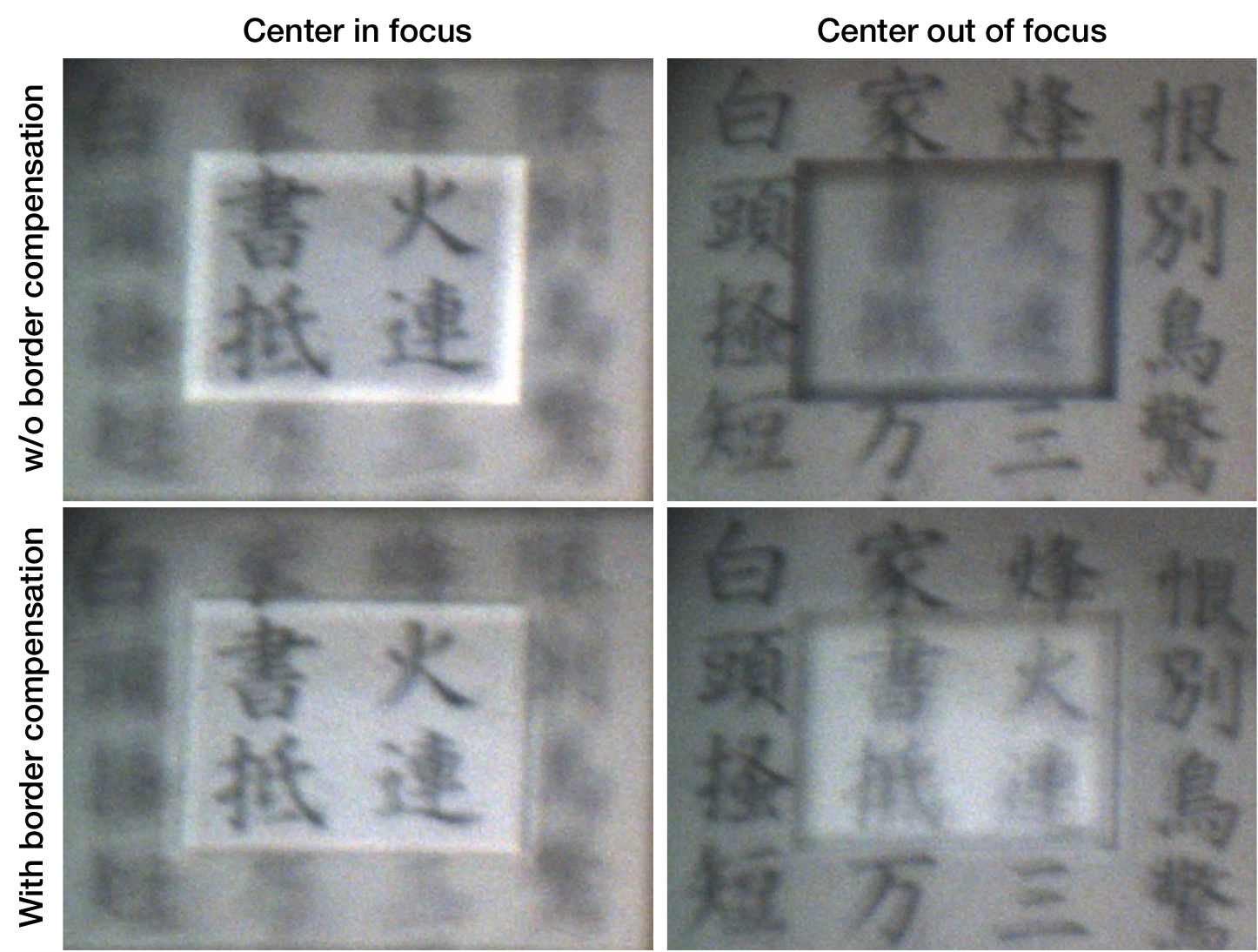} 
  \caption{Results of proposed border compensation results}
  \label{fig:Magnificationres}

\end{figure}

As proof of the effectiveness of our applied algorithm in alleviating border artifacts, we captured images using a Point Gray (Flea3 FL3-U3-32S2C) camera, equipped with a 18mm lens. We took one image before applying the algorithm and another image after border compensation (Fig. \ref{fig:Magnificationres}).

\section{Discussions}

In our study, we implemented an OST-NED system capable of modulating saliency by applying a blur effect. Our primary focus was on optically achieving this blur using an ETL. However, this approach introduced some artifacts due to magnification effects. In Section III-C, we addressed these artifacts using a straightforward yet effective algorithm.

One limitation of our current setup is that the target object must be within the acceptable DOF of the lens in front of the DMD. If the object falls outside this range, the image will appear completely blurred. In our case, we are using a 100 mm lens. This issue could be mitigated by adjusting the optical power range of the ETL, which could be dynamically manipulated based on the object's distance from the system.

Similar to other optical systems utilizing polarizers \cite{21}, the brightness of the images captured by our setup is somewhat constrained due to the use of polarizers. Additionally, the polarizers do not completely block light, resulting in some overlap between unpolarized light that passes through the system without manipulation and the manipulated light. This overlap can slightly affect the final captured image. Furthermore, precise alignment of the camera's optical center with the ETL is crucial for minimizing image distortion. Even a small misalignment can lead to artifacts, both in border compensation and in achieving a distortion-free, sharp image.

The DMD functions as a pixel-based display, requiring us to discretize the continuous range of optical powers into defined intervals. For our border compensation algorithm, we use the average optical power within each interval, which may introduce artifacts. Additionally, despite the DMD's high refresh rate, operating at 60 Hz imposes some limitations on how quickly the mask pattern can be updated. Using a DMD with a higher refresh rate and improved spatial resolution could address these limitations.

A promising direction for future work is to miniaturize the device for consumer use and incorporate an interactive blur placement feature. For the latter, we aim to align the blurred area with the scene even as the user’s head moves. To achieve this, we will apply computer vision techniques to estimate the user’s FOV during head movement, leveraging the multiple cameras integrated into current HMDs.

\section{Conclusion}
This paper presents a method for implementing saliency modulation by utilizing the optical blur effect. The main idea is to combine an ETL to manipulate the incoming light from the environment and cause blur, and a DMD to separate the areas that should be blurred from the areas that should not. We also propose an algorithm to alleviate the artifacts caused by changing magnification of the ETL. Additionally, we implemented an optical system to demonstrate the potential of our proposed method for implementing optical blur.

\section *{Disclosures}
The authors declare no conflicts of interest.

\bibliographystyle{unsrt}

\bibliography{main}

\EOD
\end{document}